\documentclass[conference]{IEEEtran}
\IEEEoverridecommandlockouts
\usepackage{cite}
\usepackage{amsmath,amssymb,amsfonts}
\usepackage{algorithmic}
\usepackage{graphicx}
\usepackage{textcomp}
\usepackage{xcolor}
\usepackage{booktabs}
\usepackage{tikz}
\usetikzlibrary{shapes.geometric, arrows.meta, positioning}

\def\BibTeX{{\rm B\kern-.05em{\sc i\kern-.025em b}\kern-.08em
    T\kern-.1667em\lower.7ex\hbox{E}\kern-.125emX}}

\begin{document}

\title{Characterizing Network Centralization and Observability in the Remote MCP Ecosystem\thanks{Accepted at the \textit{1st IEEE ICNP Workshop on Network Infrastructure and Protocols for AI Agents (NIPA 2026)}, co-located with the 34th IEEE International Conference on Network Protocols (IEEE ICNP 2026), Tempe, AZ, USA, October 2026.}}

\author{
\IEEEauthorblockN{Muhammad Abdullah Sohail}
\IEEEauthorblockA{\textit{University of Calgary} \\
Calgary, AB, Canada \\
mabdullah.sohail@ucalgary.ca}
}

\maketitle

\begin{abstract}
The Model Context Protocol (MCP) has emerged as the dominant interface for connecting autonomous agents to external data sources and execution environments. The ecosystem's transition from local process execution to remote Streamable HTTP deployments introduces unmeasured architectural and security constraints at scale. This paper presents a three-tier observability framework comprising catalog metadata ($O_0$), passive compliance signals ($O_1$), and live vulnerability analysis ($O_2$), applied to empirically characterize the public MCP server ecosystem. Evaluation of a stratified sample of 179 remote endpoints across two primary public registries reveals significant infrastructural consolidation. The Herfindahl-Hirschman Index (HHI) computed over the Autonomous System Number (ASN) distribution yields a value of 0.736, well above the 0.25 threshold for a highly concentrated market. Analysis further indicates that server authentication is strongly correlated with hosting platform choice rather than individual operator configuration, with 95\% of commercial PaaS-hosted servers enforcing gateway-level OAuth 2.1 with PKCE. The empirical results identify a Security-Observability Tradeoff observed in the current ecosystem: the platform-level authentication mechanisms that secure the majority of servers simultaneously limit automated vulnerability scanning capabilities, constraining the ability of AI gateway operators to assess tool-poisoning vectors without prior credential provisioning.
\end{abstract}

\begin{IEEEkeywords}
Model Context Protocol, AI Agents, Tool Poisoning, Network Measurement, Security, Centralization, HHI
\end{IEEEkeywords}

\section{Introduction}

The integration of Large Language Models (LLMs) with external tools has shifted the bottleneck of AI agent architectures from reasoning capabilities toward context orchestration and tool execution safety. The Model Context Protocol (MCP) \cite{mcp_spec} defines a standardized control plane in which client agents query standalone servers for context, tools, and executable prompts. Initially designed for local inter-process communication via standard I/O streams, the MCP ecosystem is transitioning toward remote deployments over Streamable HTTP to support distributed, multi-tenant agent fabrics.

This transition to externalized infrastructure exposes the MCP ecosystem to the attack surfaces of traditional web services, particularly indirect prompt injection and tool poisoning. The risk profile is compounded by the non-deterministic, autonomous nature of the client agents. A malicious or compromised remote MCP server can introduce adversarial content into an LLM's context window through manipulated schema payloads, as demonstrated by Greshake et al. \cite{greshake2023not}. Gateway operators must therefore be able to assess the security posture of a remote server before provisioning it within an active agent pipeline.

Despite this growth, no prior work has conducted a structured, server-side empirical measurement of the remote MCP deployment landscape. Prior analyses have focused primarily on local vulnerability discovery or client-side registry aggregation \cite{mcpcrawler}. This paper addresses that gap through a systematic network measurement study of the remote MCP server ecosystem. Three research questions guide the study, connected by a shared concern: whether gateway operators have sufficient visibility to make informed trust decisions about remote MCP servers.

\begin{enumerate}
    \item \textbf{Hosting consolidation} determines the blast radius of infrastructure failures. If most remote MCP servers share a single network provider, a routing disruption or platform outage could disable agent tooling across the ecosystem simultaneously.
    \item \textbf{Authentication posture} determines whether gateway operators can pre-screen a server's tool schemas for poisoning vectors before provisioning it. If authentication is enforced at the platform level rather than configured by individual operators, the observability available to external scanners depends on architectural choices outside any single developer's control.
    \item \textbf{The Security-Observability Tradeoff} ties the first two questions together: the platforms that centralize hosting also enforce authentication, which blocks the automated scanning needed to verify tool safety. Understanding this relationship is necessary for designing gateway integration workflows that remain secure under the ecosystem's current architectural constraints.
\end{enumerate}

The contributions of this work are as follows. A three-tier observability framework ($O_0$, $O_1$, $O_2$) is introduced to partition the signal space available to gateway operators across distinct phases of the server integration lifecycle. Quantitative evidence of high hosting consolidation is presented, measured by an HHI of 0.736 computed over the ASN distribution of reachable servers. The study presents empirical evidence that server authentication posture is strongly correlated with hosting platform architecture rather than individual developer configuration. The Security-Observability Tradeoff is characterized as an empirical property of the current ecosystem: the mechanism that produces the ecosystem's security baseline simultaneously reduces the observability required to verify it through automated scanning.

\section{Background and Related Work}

\subsection{The Model Context Protocol}

Developed by Anthropic and contributed to the Linux Foundation AI and Data Foundation (AAIF), MCP defines three interaction primitives between clients and servers: Tools (executable server-side functions), Resources (contextual data retrieval endpoints), and Prompts (reusable interaction templates). The June 2025 protocol specification \cite{mcp_spec} formalized Streamable HTTP as the canonical remote transport, deprecating the prior HTTP and SSE two-channel design that suffered from sticky-session constraints incompatible with stateless load balancers. The revised specification also mandated OAuth 2.1 with PKCE S256 for all hosted, multi-tenant server deployments.

\subsection{Tool Poisoning and Indirect Prompt Injection}

The threat motivating this study originates in the broader field of adversarial attacks on LLM-integrated systems. Greshake et al. \cite{greshake2023not} introduced the indirect prompt injection threat model, demonstrating that an adversary controlling content retrieved by an LLM-integrated application can subvert model behavior without direct access to the query interface. In the MCP context, this threat translates to tool poisoning: a malicious server operator embeds adversarial instructions within tool name strings, description fields, or schema annotations. When an LLM-based client enumerates available tools, these instructions are processed as legitimate context, potentially causing the agent to exfiltrate data or execute unauthorized tool calls. Empirical red-teaming of MCP infrastructure has confirmed that this vector is exploitable in production settings \cite{mcptox}. The Internet of Agents security survey \cite{ioa_survey} situates these vulnerabilities within a broader taxonomy of trust boundary violations across multi-agent systems, and analogous injection risks have been identified in the A2A protocol ecosystem \cite{a2a_security}.

\subsection{Internet Infrastructure Centralization}

Centralization of internet infrastructure is a well-documented concern in empirical network measurement research. Moura et al. \cite{moura2020clouding} quantified the concentration of global DNS query traffic across major recursive resolvers, finding that a small number of cloud providers account for a disproportionate share of resolution traffic. Complementary work on the web hosting industry \cite{moura2021hosting} documents how market dynamics drive independent operators toward shared infrastructure providers, producing emergent oligopolies at the IP layer. This body of work establishes both the methodological precedent and the analytical vocabulary for quantifying centralization in novel protocol ecosystems. The present study applies the Herfindahl-Hirschman Index to the MCP deployment layer, extending this line of measurement inquiry to AI agent tooling infrastructure.

\subsection{Ecosystem Measurement for AI Protocols}

Prior measurement studies of the AI agent ecosystem have focused on interoperability standards and transport protocol characterization \cite{mcp_survey} or congestion control mechanisms for agent traffic \cite{concur}. The closest directly comparable prior work, MCPCrawler \cite{mcpcrawler}, characterized client-side interaction patterns and aggregated marketplace metadata across several registries, finding SSE to be the dominant transport in the 2025 pre-revision dataset. The present study differs in scope: it focuses on the server-side remote deployment layer rather than clients, applies a three-tier observability framework that separates catalog-level, post-connection, and live security signals, and surfaces the Security-Observability Tradeoff, a property that client-centric analyses do not address.

\section{Measurement Methodology}

\subsection{Overview and Formal Framework}

\begin{figure}[t]
\centering
\begin{tikzpicture}[
  node distance=1.3cm and 2.8cm,
  block/.style={rectangle, draw, rounded corners, minimum height=2.4em, text centered, text width=6.5em, font=\small},
  arr/.style={-{Stealth}, thick},
  darr/.style={-{Stealth}, dashed, thick}
]
\node[block, fill=gray!20]  (t0) {Tier 0 ($O_0$)\\Catalog Metadata};
\node[block, fill=blue!15, below=of t0] (t1) {Tier 1 ($O_1$)\\Passive Probing};
\node[block, fill=red!15,  below=of t1] (t2) {Tier 2 ($O_2$)\\Live Scanning};
\node[block, fill=yellow!25, right=of t1] (gw) {Gateway Auth\\Enforcement};

\draw[arr]  (t0) -- (t1);
\draw[arr]  (t1) -- node[left, font=\scriptsize]{OPEN} (t2);
\draw[darr] (t1) -- node[above, font=\scriptsize]{AUTH\_GATED} (gw);
\draw[darr] (gw.south) -- ++(0,-0.6) -| node[above, fill=white, font=\scriptsize, pos=0.3]{Blocks $O_2$} (t2.east);
\end{tikzpicture}
\caption{The Three-Tier Observability Framework. AUTH\_GATED servers prevent Tier 2 analysis, producing the Security-Observability Tradeoff.}
\label{fig:flowchart}
\end{figure}
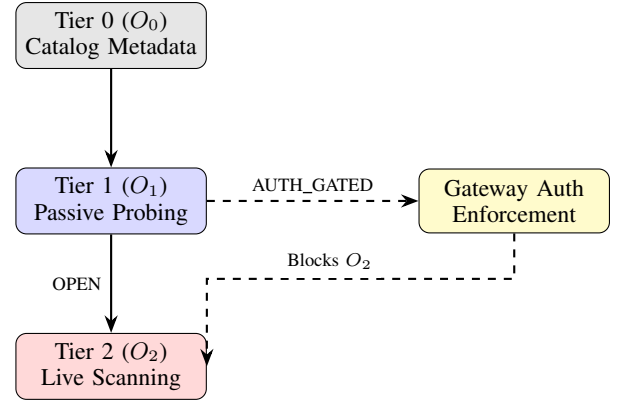

The measurement pipeline is structured around three incrementally deeper layers of observability, each corresponding to a distinct phase of what a gateway operator would encounter when integrating an external MCP server (Fig.~\ref{fig:flowchart}). Tier 0 ($O_0$) captures catalog-level metadata accessible without any network connection to the server. Tier 1 ($O_1$) captures passive signals available immediately after establishing a protocol connection. Tier 2 ($O_2$) captures active security analysis signals, available only when the server permits unauthenticated tool enumeration.

\subsection{Data Collection and Stratified Sampling}

Server endpoints were sourced from two public directories: the Anthropic-curated Official MCP Registry (\texttt{registry.modelcontextprotocol.io/v0/servers}) and Smithery, the largest commercial PaaS registry for remote MCP deployments. These two registries were selected because they are, at the time of measurement, the only publicly accessible and API-queryable directories of remote MCP server endpoints. Private, corporate, and invite-only registries are excluded from the study because they are not discoverable through public interfaces; this exclusion is itself a characteristic of the ecosystem's discoverability surface rather than a methodological limitation that can be overcome without privileged access.

To mitigate heavy-tail noise from the long-tail distribution of inactive registrations, a stratified sampling approach was applied to Smithery. Using the registry pagination API, 150 servers were collected in equal strata of 50 across three adoption tiers: the top-50 servers by \texttt{useCount} (high-adoption), servers ranked 201 to 250 (mid-tier), and servers ranked 1001 to 1050 (long-tail). This stratification ensures representation across the full adoption spectrum rather than biasing toward popular or recently-listed servers. The complete population of 29 remote-capable endpoints in the Official MCP Registry was captured as a census without stratification.

Of 179 endpoints attempted across both registries, 76 (42.5\%) accepted TCP connections within a 5000\,ms timeout. The remaining endpoints returned connection timeouts or terminal HTTP 4xx/5xx errors, consistent with servers that are registered in catalog metadata but not actively maintained in deployment. Table~\ref{tab:dataset} presents the full cross-registry breakdown.

\begin{table}[t]
\caption{Dataset Statistics: Cross-Registry Reachability and Security Posture}
\begin{center}
\begin{tabular}{lccc}
\toprule
Metric & Official Reg. & Smithery & Total \\
\midrule
Attempted       & 29    & 150   & 179  \\
Reachable       & 20    & 56    & 76   \\
Reachability Rate & 69.0\% & 37.3\% & 42.5\% \\
\midrule
\multicolumn{4}{c}{\textit{Posture of Reachable Servers}} \\
\midrule
AUTH\_GATED     & 10    & 53    & 63   \\
OPEN (SAFE)     & 7     & 0     & 7    \\
FAILED\_CONN    & 3     & 3     & 6    \\
\bottomrule
\end{tabular}
\label{tab:dataset}
\end{center}
\end{table}

\subsection{Tier 0: Catalog Metadata}

Tier 0 features were extracted from registry APIs prior to any network connection to the target servers. These features capture what a gateway operator would know about a server before establishing any network contact, representing the minimum information available for trust decisions. The $O_0$ feature set is defined as:

\begin{equation}
O_0 = \{d,\; c,\; T,\; v_s,\; \text{src}\}
\end{equation}

where $d$ is the natural language description string, $c$ is the set of categorical tags, $T$ is the declared transport type, $v_s$ is the declared protocol version, and $\text{src}$ denotes the source registry. Popularity proxies such as \texttt{useCount} and the binary \texttt{verified} badge were excluded to ensure the framework measures intrinsic server characteristics rather than platform-assigned endorsement signals.

\subsection{Tier 1: Post-Connection Passive Probing}

For the 76 reachable servers, the measurement probe initiated protocol-compliant Streamable HTTP handshakes, issuing the \texttt{initialize} method request followed by the \texttt{tools/list} request. The $O_1$ feature set is defined as:

\begin{equation}
O_1 = \{k,\; \bar{l},\; \bar{q},\; \text{tls},\; pv\}
\end{equation}

where $k$ is the tool count returned by \texttt{tools/list}, $\bar{l}$ is the mean description length across exposed tools, $\bar{q}$ is the mean input schema quality score (fraction of parameters carrying both a \texttt{type} declaration and a \texttt{description} string), \texttt{tls} is a binary indicator of HTTPS enforcement, and $pv$ is the negotiated protocol version. A composite compliance score $\sigma$ was computed per server to summarize adherence to specification recommendations in a single scalar, enabling comparison across servers:

\begin{equation}
\sigma = \frac{1}{3}\Bigl(\text{tls} + \mathbb{1}[pv \geq \texttt{2024-11-05}] + \bar{q}\Bigr)
\end{equation}

$O_1$ features are unavailable for AUTH\_GATED servers, as the authentication challenge is issued prior to MCP handshake completion.

\subsection{Tier 2: Live Security Analysis}

Ground-truth security labels were established using the public \texttt{agentseal scan-mcp --url} CLI \cite{agentseal}. This pipeline evaluates tool schemas against a curated repository of prompt-injection patterns and semantic poisoning heuristics, applying embedding cosine-distance calculations to detect semantically obfuscated payloads. Each server receives a trust score $\tau \in [0, 100]$, binarized as \texttt{SAFE} when $\tau \geq 70$ and \texttt{NEEDS\_REVIEW} otherwise. The parser was extended to classify servers returning HTTP 401 or 403 responses as \texttt{AUTH\_GATED}, preventing conflation of authentication enforcement with tool payload safety.

\section{Hosting Consolidation}

\subsection{Cross-Registry Infrastructure Analysis}

\begin{figure}[t]
\centerline{\includegraphics[width=\columnwidth]{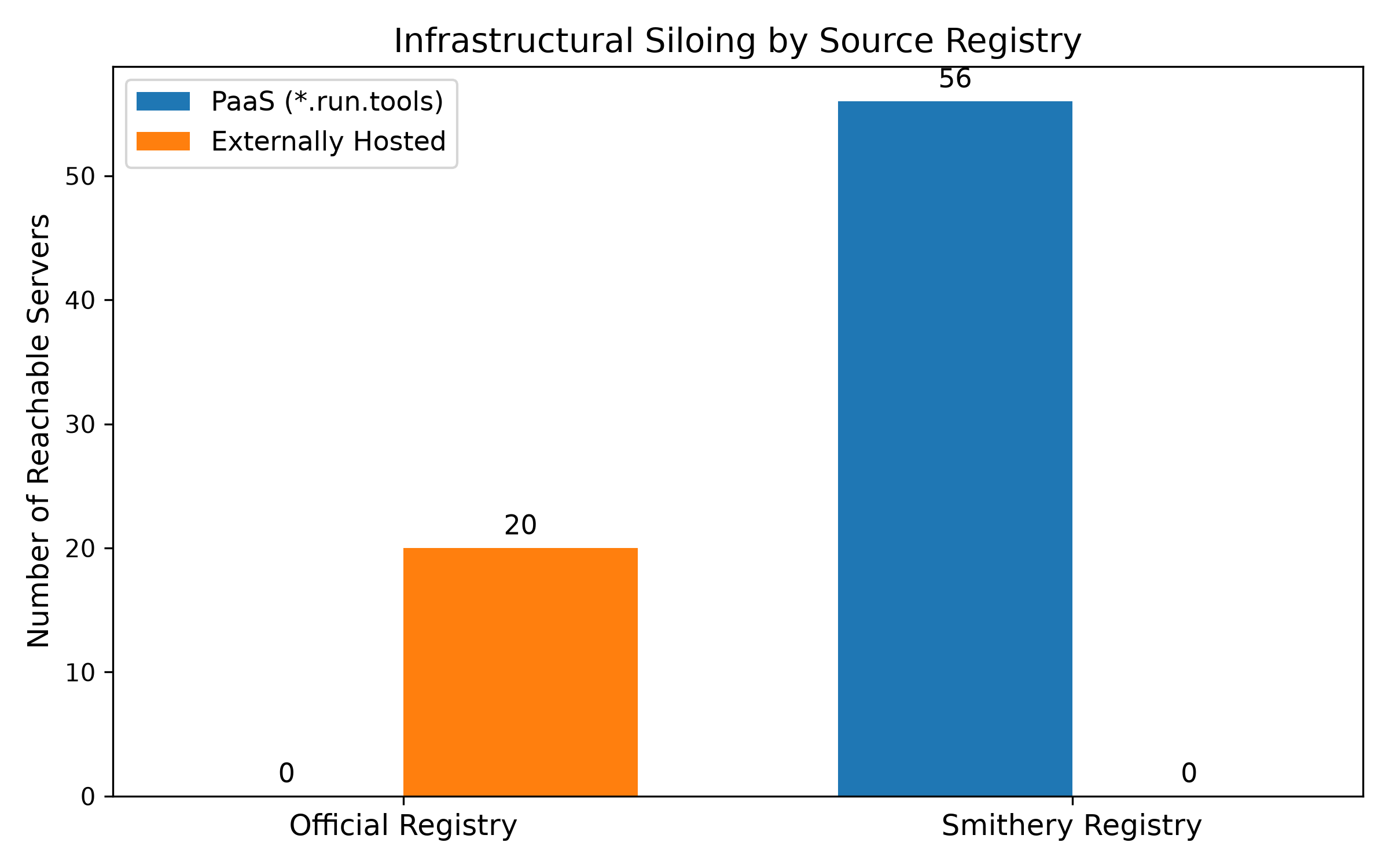}}
\caption{Hosting distribution by source registry. Zero-count annotations highlight complete absence of cross-pollination.}
\label{fig:consolidation}
\end{figure}

The first research question examined the degree of infrastructural concentration across the remote MCP ecosystem. Of the 76 reachable endpoints, 56 (73.7\%) are hosted on Smithery's \texttt{*.run.tools} domain. The aggregate figure, however, is partially an artifact of the 150:29 registry sampling ratio. A cross-registry comparison removes this confound: among Official MCP Registry servers, sampled independently of Smithery, 0 of 20 resolve to \texttt{*.run.tools}. Among Smithery-registry servers, 56 of 56 reachable endpoints resolve exclusively to \texttt{*.run.tools}. There is complete absence of cross-registry hosting. No Official Registry server was deployed on Smithery PaaS infrastructure, and no Smithery-listed server was hosted on a third-party provider (Fig.~\ref{fig:consolidation}).

\subsection{ASN-Level Concentration}

\begin{figure}[t]
\centerline{\includegraphics[width=\columnwidth]{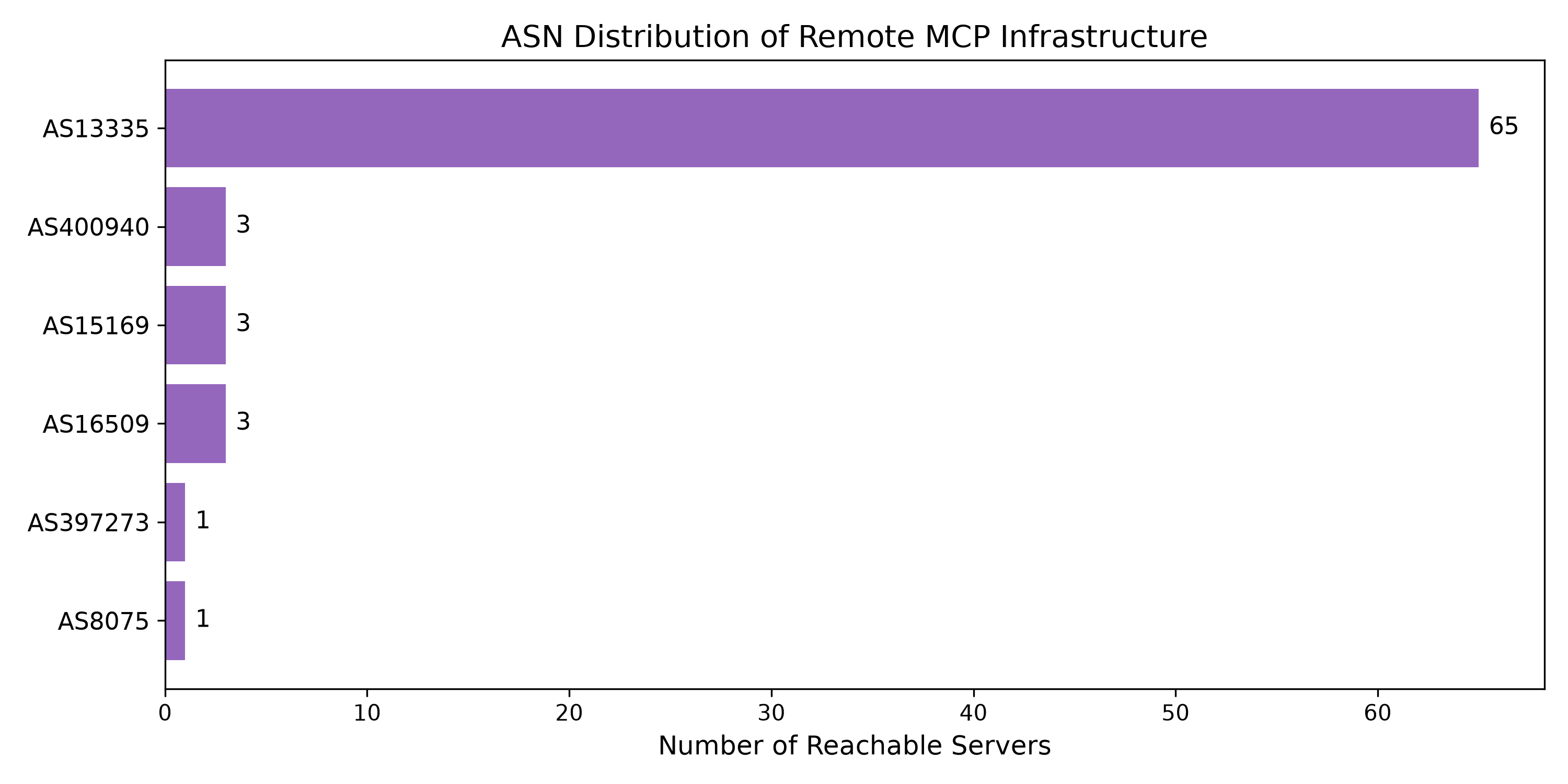}}
\caption{ASN distribution of 76 reachable endpoints. AS13335 (Cloudflare) accounts for 85.5\% of servers.}
\label{fig:asn}
\end{figure}

To evaluate centralization at the network layer, ASN records were resolved for each reachable endpoint via the \texttt{ip-api.com} enrichment API. Fig.~\ref{fig:asn} presents the resulting distribution. The dominant ASN is AS13335 (Cloudflare, Inc.), which accounts for 65 of 76 servers (85.5\%). The remaining servers are distributed across AS400940 (3 servers), AS15169 (Google Cloud, 3 servers), AS16509 (Amazon Web Services, 3 servers), AS397273 (1 server), and AS8075 (Microsoft Azure, 1 server).

To formally quantify this concentration, the Herfindahl-Hirschman Index (HHI) was computed. HHI is the standard metric used in antitrust economics and in prior internet centralization studies \cite{moura2020clouding} for quantifying market concentration from share distributions. It is defined as $\sum_i s_i^2$ where $s_i$ is the fractional share of entity $i$. Values below 0.15 indicate an unconcentrated market, values between 0.15 and 0.25 indicate moderate concentration, and values above 0.25 indicate high concentration under DOJ antitrust guidelines. For the observed ASN distribution:

\begin{align}
\text{HHI} &= 0.855^2 + 0.039^2 + 0.039^2 \nonumber \\
           &\quad + 0.039^2 + 0.013^2 + 0.013^2 = 0.736
\end{align}

The observed HHI of 0.736 is 2.94 times the high-concentration classification threshold. By this metric, the remote MCP ecosystem is more concentrated at the network layer than most markets that have attracted regulatory antitrust scrutiny. This finding parallels Moura et al.'s documentation of DNS traffic centralization \cite{moura2020clouding} and extends the same dynamic to AI agent tooling infrastructure.

\section{Authentication as a Platform Property}

\subsection{Posture Distribution by Hosting Type}

\begin{figure}[t]
\centerline{\includegraphics[width=\columnwidth]{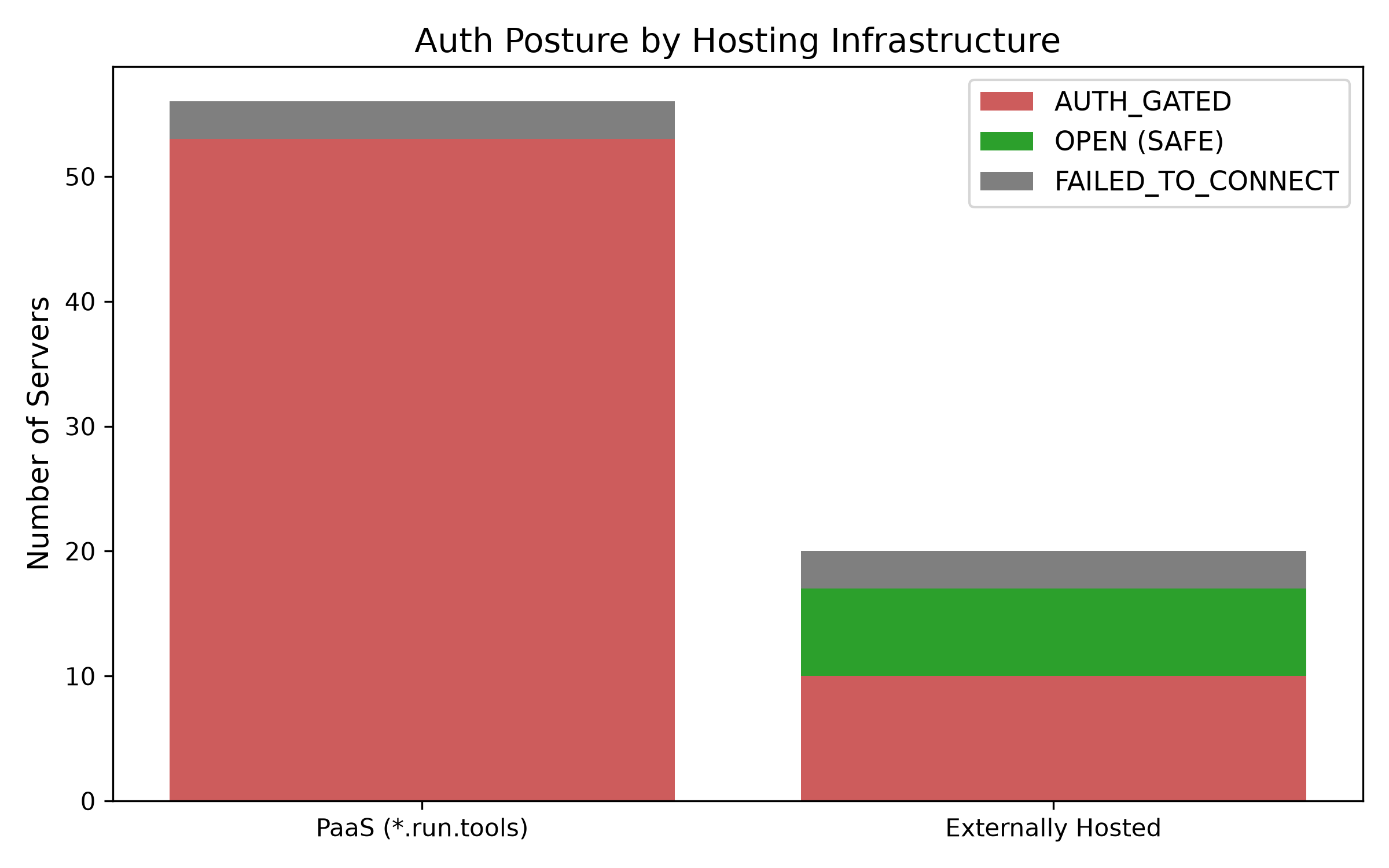}}
\caption{Security posture distribution stratified by hosting infrastructure type.}
\label{fig:auth}
\end{figure}

The second research question examined whether authentication enforcement correlates with hosting platform choice or varies independently across individual operators. Tier 2 ($O_2$) data across the 76 reachable servers reveals a clear pattern (Fig.~\ref{fig:auth}). Of the 56 Smithery-hosted servers, 53 (94.6\%) returned \texttt{AUTH\_GATED} responses, zero permitted unauthenticated tool enumeration, and 3 returned transport-level failures. Of the 20 externally-hosted servers, 10 (50.0\%) were \texttt{AUTH\_GATED} and 7 (35.0\%) permitted open enumeration.

The conditional probability $P(\texttt{AUTH\_GATED} \mid \texttt{Smithery}) = 0.946$ compared to $P(\texttt{AUTH\_GATED} \mid \texttt{External}) = 0.500$. The 44.6 percentage-point differential is consistent with authentication being strongly associated with platform hosting choice. However, the cross-sectional design of this study cannot establish causation: operators who self-select into Smithery may also be more likely to configure authentication independently, and the observed correlation could reflect both platform defaults and operator demographics.

\subsection{OAuth 2.1 at the Gateway Layer}

The June 2025 MCP specification mandates OAuth 2.1 with PKCE S256 for all hosted remote deployments, replacing the prior bearer-token model \cite{mcp_spec}. Smithery implements this mandate at the gateway layer: the \texttt{*.run.tools} domain fronts all hosted servers behind a centralized OAuth authorization server, requiring access tokens for MCP session initialization. From the perspective of an external measurement probe, this architecture is operationally indistinguishable from a developer-configured authentication decision. The finding that zero Smithery servers permit unauthenticated access, compared to 35\% of external servers, is consistent with authentication being a consequence of hosting platform defaults rather than individual operator security practice.

\subsection{Compliance Characteristics of the Open Population}

Granular $O_1$ compliance data was collected for the seven externally-hosted open servers. Tool count $k$ ranged from 0 to 13 (mean 7.3, median 6), confirming that the \texttt{SAFE} classifications assigned by the Tier 2 pipeline are not vacuous: these servers expose substantive tool interfaces. Mean schema quality $\bar{q}$ was 0.61, indicating that a majority of exposed parameters carried type declarations but that description-string completeness was mixed. All seven open servers enforced HTTPS transport. Mean composite compliance $\bar{\sigma}$ was 0.74, indicating adequate but incomplete adherence to specification recommendations.

\section{The Security-Observability Tradeoff}

\subsection{Classifier Design}

The third research question examined whether $O_0$ catalog metadata and $O_1$ passive compliance signals could predict $O_2$ security vulnerability classifications, providing a passive pre-screening capability usable within the authentication-as-platform constraint. A Logistic Regression classifier was trained to predict the AgentSeal vulnerability verdict (\texttt{SAFE} vs. \texttt{NEEDS\_REVIEW}). The $O_0$ description string was encoded using the \texttt{all-MiniLM-L6-v2} sentence-transformer model to produce a 384-dimensional semantic embedding, concatenated with the five scalar $O_1$ features to yield a 389-dimensional input vector. L2 regularization with $C = 1.0$ was applied.

\subsection{Single-Class Degeneracy}

Classification was constrained to the seven open servers for which both $O_1$ and $O_2$ data were available. All seven received a \texttt{SAFE} verdict from the AgentSeal scanner, meaning no concrete tool-poisoning payloads were detected in any of the scannable servers. The target label vector exhibits zero variance, rendering the classification task ill-defined: a zero-rule classifier achieves 100\% accuracy by predicting \texttt{SAFE} unconditionally. We acknowledge that drawing statistical conclusions from seven observations is not possible with conventional significance testing. However, the result is informative precisely because it is structurally determined: the population accessible to unauthenticated scanning is small not by sampling choice but because platform-level authentication gates the largest server population, preventing their inclusion in any live scanning dataset. The single-class outcome is therefore an empirical consequence of the ecosystem's architecture rather than a failure of experimental design.

\subsection{Formal Characterization}

Let $S$ denote the full population of remotely reachable MCP servers. Let $S_{\text{auth}} \subset S$ denote the subset for which platform-level authentication is enforced, and let $S_{\text{open}} = S \setminus S_{\text{auth}}$ be the complement. The Tier 2 security analysis function $f : S_{\text{open}} \to \{\texttt{SAFE}, \texttt{NEEDS\_REVIEW}\}$ is defined only over $S_{\text{open}}$.

Empirically, $|S_{\text{auth}}| / |S| = 63/76 = 0.829$. In the current ecosystem state, any mechanism increasing $|S_{\text{auth}}| / |S|$ reduces $|S_{\text{open}}| / |S|$, shrinking the domain of $f$. The data shows this tradeoff is already in a pronounced state: 82.9\% of the reachable ecosystem is opaque to Tier 2 analysis. This characterization reflects the ecosystem as measured in June 2025 and should not be interpreted as a universal property of all possible MCP deployment architectures; alternative designs (such as the transparency log mechanism proposed in Section~\ref{sec:governance}) could decouple authentication from observability.

\section{Discussion}

\subsection{Centralization as a Systemic Risk}

The HHI of 0.736, driven by a single-ASN dominance of 85.5\%, introduces systemic fragility that is independent of any individual server's security posture. A routing disruption, BGP prefix hijack, or platform-level incident affecting the dominant ASN would degrade or compromise connectivity for the majority of the remote MCP ecosystem simultaneously. This dependency structure is analogous to the single-point-of-failure vulnerability documented in centralized DNS resolver deployments \cite{moura2020clouding}, with the additional consequence that the dependent entities are autonomous agents rather than passive query resolvers.

\subsection{Design Implications for Gateway Operators}

The Security-Observability Tradeoff has direct operational consequences for developers of AI gateways and agent routing fabrics. Given that $O_2$ active scanning is structurally unavailable for 82.9\% of the reachable ecosystem in our dataset, gateway operators should design integration workflows that require credential provisioning before any security assessment phase. An \texttt{AUTH\_GATED} classification confirms platform-level authentication enforcement but provides no information about the safety of the tool schemas protected behind that authentication layer. Practical design responses include maintaining a local trust ledger of $O_0$ and $O_1$ signals aggregated during the integration handshake, requiring server operators to provide a signed tool manifest at onboarding time, and applying differential trust policies for PaaS-hosted servers (where platform-level audit commitments may exist contractually) relative to externally-hosted servers, where the 50\% authentication adoption rate observed here indicates higher variance in operator security practice.

\subsection{Protocol-Level Governance}
\label{sec:governance}

The following protocol extensions are presented as design directions motivated by the measurement results, not as validated solutions. They would require further specification work and community review before adoption.

The Linux Foundation AAIF, which governs the MCP specification, could consider two extensions in a future revision. The first is a Tool Schema Transparency Log: a cryptographically-signed, append-only log of tool schema versions, structurally analogous to the Certificate Transparency log infrastructure for X.509 certificates. Under this model, a server's tool schemas would be submitted to a public log at deployment time, producing a signed tree head verifiable by gateway operators against the schema hash returned during the OAuth handshake. This mechanism would preserve $O_2$-equivalent security signals in the presence of authentication, without requiring unauthenticated tool enumeration. The second is a set of standardized security metadata fields at the $O_0$ catalog level, specifically \texttt{schema\_hash}, \texttt{scan\_attestation\_uri}, and \texttt{last\_scan\_timestamp}, surfaced in registry API responses. These fields would permit registry-level security characterization prior to any protocol-level connection, addressing the measurement gap exposed by the single-class degeneracy without requiring changes to the server-to-client wire protocol.

\subsection{Limitations}

This study has several limitations. The dataset is a cross-sectional measurement at a single point in time (June 2025). Given the rapid evolution of the MCP ecosystem, the hosting distribution, authentication rates, and registry composition may change substantially in subsequent months. Longitudinal measurement would strengthen the empirical support for these findings. The sample is drawn from two public registries; enterprise-internal, invite-only, and privately hosted MCP servers are not represented. The Tier 2 analysis is constrained to seven open servers, which is too small to support statistical generalization about tool-poisoning prevalence across the broader ecosystem. The correlation between hosting platform and authentication posture, while strong, cannot be interpreted as causal without a controlled experiment or longitudinal design that observes operator behavior before and after platform migration.

\section{Ethics}

This measurement study was conducted in accordance with responsible disclosure and ethical network measurement practice. The experimental pipeline used deterministic protocol handshakes and did not invoke the \texttt{tools/call} endpoint or execute any server-side tool at any point. The crawler was rate-limited to a maximum of three concurrent connections with a one-second inter-request delay per target, to minimize operational impact on measured infrastructure. All active vulnerability assessments were performed using the public \texttt{agentseal} CLI as intended by the vendor, targeting only publicly reachable endpoints listed in public registries. No credentials or authentication tokens belonging to third parties were used or solicited.

\section{Conclusion}

This paper presents a systematic server-side measurement of the remote Model Context Protocol ecosystem. A three-tier observability framework ($O_0$, $O_1$, $O_2$) was applied to a stratified sample of 179 endpoints drawn from two primary public registries. The empirical results show high hosting consolidation at the network layer, measured by an HHI of 0.736, with a single Autonomous System (AS13335, Cloudflare) hosting 85.5\% of reachable servers. Authentication enforcement is strongly correlated with hosting platform choice, with 94.6\% of commercial PaaS-hosted servers enforcing OAuth 2.1 with PKCE at the gateway layer. The Security-Observability Tradeoff, as observed in this dataset, shows that 82.9\% of the reachable ecosystem is opaque to automated security analysis, and that increases in platform-level authentication adoption reduce the feasible domain of Tier 2 measurement. Future work should include longitudinal measurement to track how these properties evolve, and should explore cryptographic transparency log mechanisms or standardized attestation metadata at the catalog layer to decouple authentication from observability.

\bibliographystyle{IEEEtran}
\bibliography{references}

\end{document}